\documentclass[pra,aps,showpacs,floatfix,twocolumn]{revtex4-2}
\usepackage{amsmath}
\usepackage{amsfonts}
\usepackage{amssymb}
\usepackage{revsymb}
\usepackage{graphicx}
\usepackage{physics}
\usepackage{mathtools}
\usepackage{tabularx}
\usepackage{booktabs}
\usepackage[table]{xcolor}

\newcommand{\fref}[1]{Fig.~\ref{#1}}
\newcommand{\tref}[1]{Table~\ref{#1}}
\newcommand{\Lu}{$^{176}\mathrm{Lu}^+$}

\begin{document}

\title{Land\'{e} g-factor measurements for the 5d6s \textsuperscript{3}D\textsubscript{2} hyperfine levels of \textsuperscript{176}Lu\textsuperscript{+}}

\author{Qi Zhao}
\affiliation{Centre for Quantum Technologies, National University of Singapore, 3 Science Drive 2, 117543 Singapore}
\author{M. D. K. Lee}
\affiliation{Centre for Quantum Technologies, National University of Singapore, 3 Science Drive 2, 117543 Singapore}
\author{Qin Qichen}
\affiliation{Centre for Quantum Technologies, National University of Singapore, 3 Science Drive 2, 117543 Singapore}
\author{Zhao Zhang}
\affiliation{Centre for Quantum Technologies, National University of Singapore, 3 Science Drive 2, 117543 Singapore}
\author{N. Jayjong}
\affiliation{Centre for Quantum Technologies, National University of Singapore, 3 Science Drive 2, 117543 Singapore}
\author{K. J. Arnold}
\affiliation{Centre for Quantum Technologies, National University of Singapore, 3 Science Drive 2, 117543 Singapore}
\affiliation{Temasek Laboratories, National University of Singapore, 5A Engineering Drive 1, Singapore 117411, Singapore}
\author{M. D. Barrett}
\affiliation{Centre for Quantum Technologies, National University of Singapore, 3 Science Drive 2, 117543 Singapore}
\affiliation{Department of Physics, National University of Singapore, 2 Science Drive 3, 117551 Singapore}
\begin{abstract}
We report measurements of the Land\'{e} g-factors for the 5d6s \textsuperscript{3}D\textsubscript{2} hyperfine levels of \textsuperscript{176}Lu\textsuperscript{+} to a fractional inaccuracy of $5\times 10^{-7}$.  Combining these measurements with theoretical calculations allows us to estimate hyperfine-mediated modifications to the quadrupole moments for each state and infer a value of $\delta\Theta = 1.59(34)\times 10^{-4} \,ea_0^2$ for the residual quadrupole moment of the $^1S_0\leftrightarrow{^3}D_2$ hyperfine-averaged clock transition.  
\end{abstract}

\maketitle

\section{Introduction}
Singly ionized lutetium supports two long lived clock states (\textsuperscript{3}D\textsubscript{1} and \textsuperscript{3}D\textsubscript{2}).    In both cases, averaging over all hyperfine levels of a fixed $m_F$ highly suppresses systematics associated with the electronic angular momentum $J$, which renders an effective $J=0$ level \cite{barrett2015developing}.  The corresponding $^1S_0\leftrightarrow{^3}D_J$ hyperfine-averaged clock transitions provide frequency references with exceptionally low systematics \cite{zhiqiang2023176lu+,arnold2024validating}.  In \cite{arnold2024validating} we have argued that the ratio of the two frequency references measured within the one system would provide an independent metric to validate system performance.  However, systematics for the $^1S_0\leftrightarrow{^3}D_2$ transition are yet to be accurately quantified.

In this work, we investigate the residual quadrupole moment (RQM) for the $^1S_0\leftrightarrow{^3}D_2$ transition, which arises from hyperfine-mediated coupling between fine-structure levels \cite{beloy2017hyperfine,zhiqiang2020hyperfine}.  For $J=0$ levels, the result is that the quadrupole moment is not strictly zero as expected for states with $J\leq1/2$ \cite{beloy2017hyperfine}.  For the same reason, cancellation of quadrupole shifts by hyperfine-averaging is imperfect but the remaining shift is still described by a small non-zero RQM.  As discussed in \cite{zhiqiang2020hyperfine}, the effect is larger for lutetium as it arises from the more dominant coupling to the nuclear magnetic moment, which is absent for true $J=0$ levels.  However, the same coupling influences g-factors of the hyperfine levels.  As demonstrated for the $^1S_0\leftrightarrow{^3}D_1$ transition \cite{zhiqiang2020hyperfine}, this allows estimates for the RQM to be extracted from precision g-factor measurements together with theoretical calculations.  For the $^1S_0\leftrightarrow{^3}D_2$ transition, this approach is complicated by the larger number of couplings involved.  Nevertheless we are able to provide an estimate that bounds the clock shift to the low $10^{-19}$ level for typical operating conditions.

In Sect.\ref{Sect:experiment} we give a description of the experimental system used to carry out the measurements summarized in Sect.\ref{Sect:measurements}.  In Sect.\ref{Sect:analysis} we detail the analysis and approximations made to infer the RQM.

\section{Experimental system}
\label{Sect:experiment}
Experiments are carried out in a linear Paul trap similar in design to those used in previous work~\cite{zhiqiang2023176lu+}, and specifically is the trap labeled Lu-1 in~\cite{arnold2024enhanced}. It consists of two axial end caps separated by $\sim 2.7$\,mm and four rods arranged on a square with sides 1.2\,mm in length. All electrodes are made from $\sim 0.47$ mm electropolished copper–beryllium rods. The radio-frequency (rf) potential at a frequency of $\Omega_\mathrm{rf}=2\pi \times 21.7\,$MHz is delivered via a quarter-wave helical resonator. Together with static potentials of 9\,V on the end caps and 0.5\,V on two of the diagonally opposite
rods, the measured trap frequencies for a single \Lu are $\sim 2\pi \times (648,614,177)$ kHz, with the lowest trap frequency along the trap axis, defined as $\hat{\textbf{z}}$. A dc magnetic field of $\sim 2.57$\,mT is applied in the $xz$-plane at an angle $\phi = 37^{\circ}$ with respect to $\hat{\textbf{x}}$, which defines the quantization axis.  Although we note that the angle of the field is not particularly relevant for this work.

\begin{figure}[h]
\begin{center}
\includegraphics[width=1.0\linewidth]{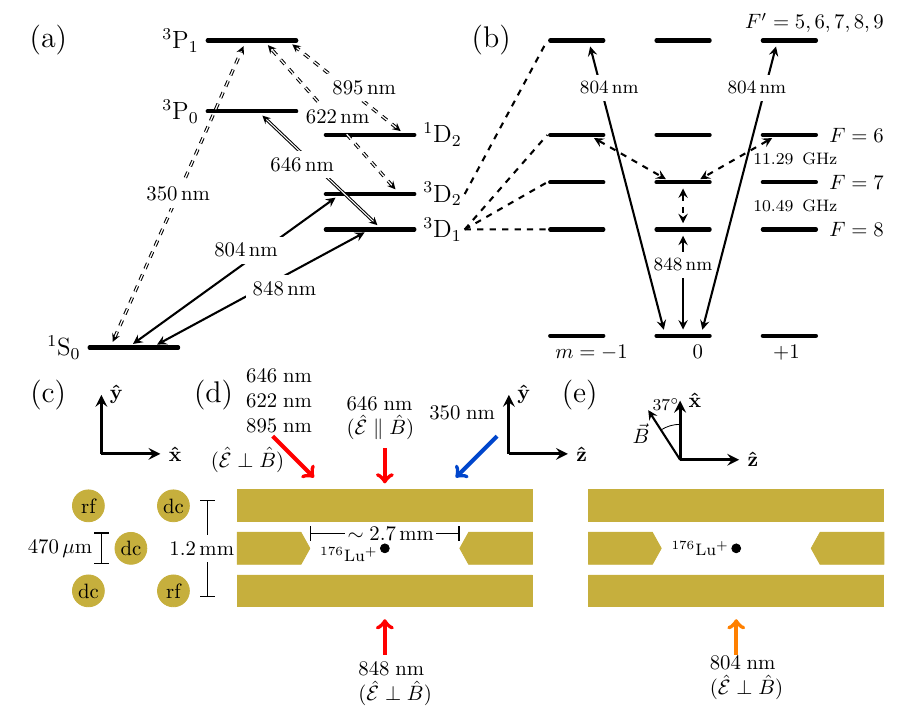}
\caption{\label{fig:schematic} (a) Atomic-level structure of $^{176}$Lu$^+$ showing the wavelengths of repump (double dashed line), cooling/detection (double line), and clock (solid line) transitions used.  (b) Optical (solid line) and microwave (dashed line) clock transitions used in the probe sequence. (c-e) Ion trap geometry with orientation and relevant polarisations of all lasers.  As the 350\,nm repump laser couples $F\rightarrow F+1$ the polarisation is not important.}
\end{center}
\end{figure}

The relevant level structure of \Lu is shown in Fig \ref{fig:schematic}. Lasers at 350, 622 and 895 nm initialize population in the $^3D_1$ state. A laser at 646 nm, which has three independent frequency components coupling $^3D_1(F=6,7,8)$ to $^3P_0(F=7)$, provides Doppler cooling and state detection for the $^3D_1$ state with fluorescence collected onto a single photon counting module (SPCM).  Switching the $^3D_1(F=7)$ to $^3P_0(F=7)$ component with the $\pi$-polarized 646-nm beam facilitates state preparation by optical pumping into $\ket{^3D_1,7,0}$. Lasers at 848 nm and 804 nm drive the $^1S_0 \leftrightarrow {^3}D_1$ and $^1S_0 \leftrightarrow {^3}D_2$ clock transitions respectively. All laser beams are switched with acousto-optic modulators. Laser configurations relative to the
trap are as shown in \fref{fig:schematic}. Two microwave antennas are used to drive the $\Delta m=0,\pm 1$ microwave transitions between hyperfine levels. The antenna driving $\ket{^3D_1,7,0} \leftrightarrow  \ket{^3D_1,6,\pm1}$ transitions is manually positioned to give approximately equal coupling to the $\Delta m=\pm 1$ transitions and reduced coupling for $\Delta m=0$.

The 848-nm clock laser is locked to a 10 cm long ultralow expansion (ULE) cavity with finesse of $\sim 4\times 10^5$ and has a line-width of $\sim 1$ Hz. The 804-nm laser is phase-locked to an optical frequency comb (OFC), which is itself phase-locked to the 848-nm laser. Periodic interrogation of the $\ket{^1S_0,7,0} \leftrightarrow \ket{^3D_1,8,0}$ clock transition is performed to compensate the slow drift of the ULE cavity and ensure both 804 and 848 nm lasers remain resonant with their respective optical transitions. All rf and microwave sources are referenced to a hydrogen maser.

\section{Measurements}
\label{Sect:measurements}
The Land\'e g-factors of the \Lu~$^3D_1$($F=6$) and $^3D_2$($F=5,6,7,8,9$) states are denoted $\bar{g}_6$ and $g_F$ respectively. Using a single ion, we measure the Zeeman splittings between $m=\pm1$ magnetic substates in the $^3D_1$($F=6$) and $^3D_2(F)$ manifolds to evaluate the ratios $r_{F} \equiv g_F/\bar{g}_6$. The $g_F$ for $^3D_2$ are then inferred using the previously reported value of $\bar{g}_6$~\cite{zhiqiang2020hyperfine}.

The experiment consists of interleaved measurements of $\ket{^3D_1,6,\pm 1}$ and $\ket{^3D_2,F,\pm 1}$ Zeeman splittings. Both measurement sequences start with a conditional preparation in the $^3D_1$ state. A 20 ms pulse including repump lasers and the 646 nm cooling laser at increased detuning and intensity is applied followed by a 5 ms state detection pulse during which the counts on the SPCM are recorded. This is repeated until the counts recorded in the 5\,ms interval exceed a fixed threshold. This ensures that experiment does not proceed if the ion is either (i) not in $^3D_1$ state or (ii) in a highly excited thermal state (e.g. due to background gas collision) such that the fluorescence rate is reduced.

The sequence for interrogating the $\ket{^3D_1,6,\pm 1}$ Zeeman states begins with a 5 ms Doppler cooling pulse followed by 1.5 ms of optical pumping to $\ket{^3D_1,7,0}$ with $>95\%$ fidelity. Rabi spectroscopy of the $\ket{^3D_1,7,0}\leftrightarrow \ket{^3D_1,6,\pm 1}$ transition is performed with a $\tau_6=6\,\mathrm{ms}$ duration microwave pulse. Lastly, the remaining $\ket{^3D_1,7,0}$ population is shelved to $\ket{^1S_0,7,0}$ via $\ket{^3D_1,8,0}$ in a two step process of sequential microwave (4 ms) and optical (3.4 ms) $\pi$-pulses, before ${^3}D_1$ state detection with 646\,nm light.  State detection is performed with a real time adaptive Bayesian algorithm and requires $\lesssim 1$ ms \cite{myerson2008high}. 

The sequence for interrogating the $\ket{^3D_2,F,\pm 1}$ Zeeman states begin with conditional $\ket{^1S_0,7,0}$ state preparation. This is achieved by repeating the sequence of Doppler cooling pulse, optical pumping to $\ket{^3D_1,7,0}$, the two step coherent transfer to $\ket{^1S_0,7,0}$ described above, and ${^3}D_1$ state detection until a `dark' detection result heralds success. In this way $>$99\% state preparation is achieved. Rabi spectroscopy of the $\ket{^1S_0,7,0}$ to $\ket{^3D_2,F,\pm 1}$ transitions, for $F=5,...,9$, is performed on the 804 nm optical transition. The 804 nm interrogation time is set to $\tau_{F} \approx  \frac{\bar{g}_6}{g_{F}}\tau_6$ so that the sensitivity to magnetic field fluctuations is approximately the same for all Zeeman pairs. Finally, remaining $\ket{^1S_0,7,0}$ population is shelved back to $\ket{^3D_1,8,0}$ with an 848 nm $\pi$-pulse, followed by ${^3}D_1$ state detection.

\begin{figure}
\centering
\includegraphics[width=1.0\linewidth]{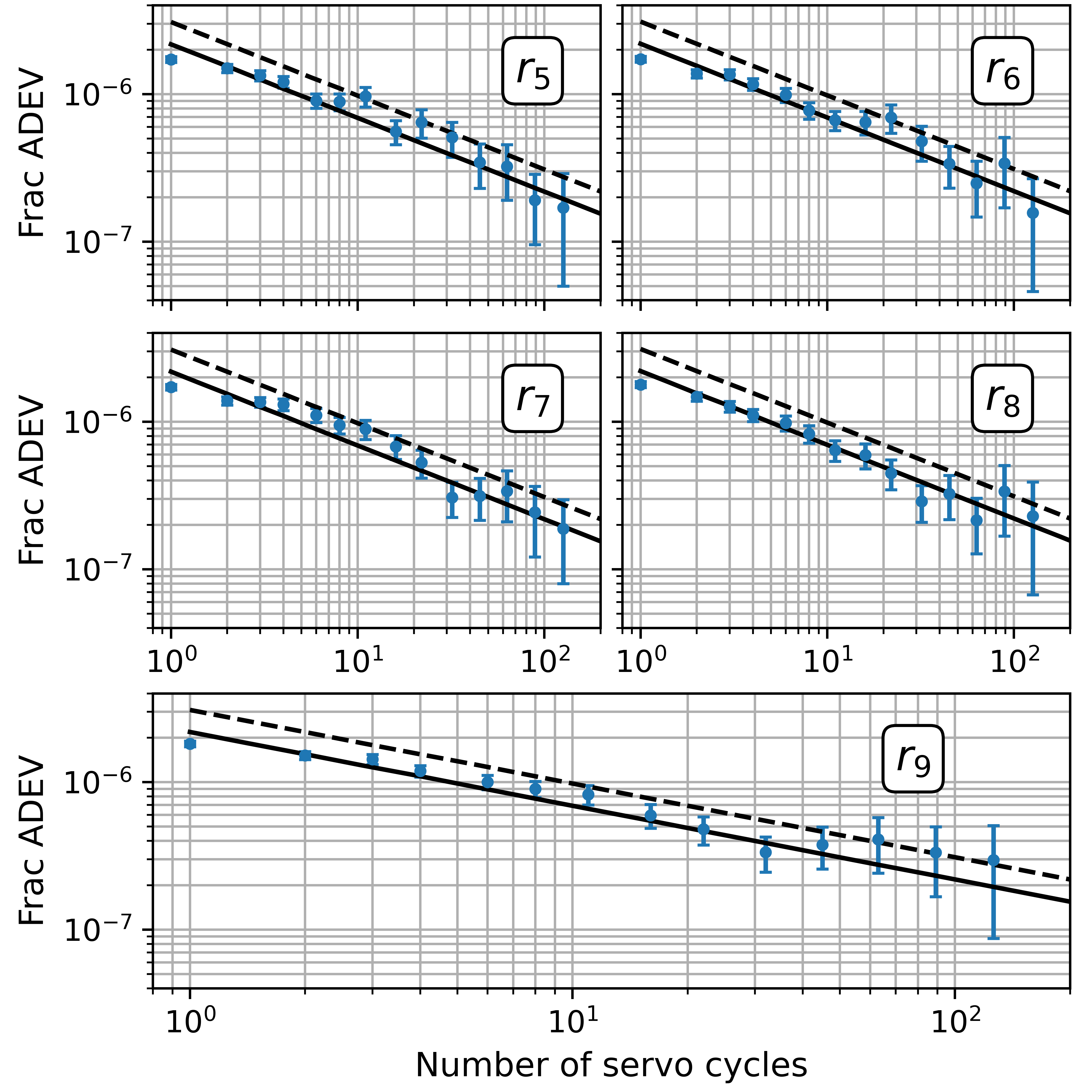}

\caption{\label{fig:allan} Fractional Allan deviation of the ratios $r_F=g_F/\bar{g}_6$. Solid black lines indicates the QPN limit, and dashed lines a factor of $\sqrt{2}$ above the QPN limit.}
\end{figure}

One cycle of the interleaved g-factor ratio measurement consists of 8 sequences for Rabi interrogation at the half maximum on both sides of the $m=\pm1$ Zeeman states for both $^3D_1$($F=6$) and $^3D_2$($F$) transitions. Every 30 cycles an integrating servo is updated to track the Zeeman splittings for both $^3D_1$ and $^3D_2$.  Data was collected for approximately 3 hours for each $F$ level and the Allan deviations of the resulting ratios $r_{F}$ are shown in \fref{fig:allan}.

At the quantum projection noise (QPN) limit, the Allan deviation for the ratio of Zeeman splittings between two levels can be written as $\sigma_0/\sqrt{M}$, where $M$ is the number of servo updates and $\sigma_0$ is the fractional resolution from a single update. For Rabi spectroscopy used here, this can be written as~\cite{zhiqiang2020hyperfine}
\begin{equation}
\sigma_0 = \frac{1.656 \hbar}{2\mu_B B \sqrt{N}}\left( \frac{1}{(g_1 \tau_1)^2}+\frac{1}{(g_2 \tau_2)^2} \right)^{1/2}
\end{equation}
where $\mu_B$ is the Bohr magneton, $B$ is the applied magnetic field, $N$ is the number of interrogations per side of a given transition, $g_k$ are the g-factors for levels involved, and $\tau_k$ are the respective interrogation times. Here $g_k\tau_k$ was the same for all servos, with a corresponding measurement resolution $\sigma_0 \approx 2\times10^{-6}$. The Allan deviations are observed to be slightly elevated above the QPN, which is attributed to the magnetic field noise which is comparable to the projection noise. To account for this, we take the statistical uncertainty in the result to be $\sqrt{2}$ above the projection noise limit, as indicated by the dashed lines in \fref{fig:allan}. 

The leading systematic effect is due to off-resonant couplings to other $\ket{^1S_0,7,m} \leftrightarrow \ket{^3D_2,F,m=\pm 1}$ optical transitions. To evaluate these shifts, we measured the relative coupling strength on 804 nm $\Delta m = (0,\pm 1, \pm 2)$ transitions both before and after each g-factor measurement. For the experimentally measured $\pi$-times used to drive the $\ket{^3D_2, F, \pm 1}$ lines, we estimate the shifts as given \tref{tab:probestark}.

\begin{table}[h]
    \caption{\label{tab:probestark}
    Values and uncertainties for $r_F$ including correction for systematic effects: ($xx$) indicates the uncertainty for a given quantity, and $[-E]$ indicates a power of 10 ($\times 10^{-E}$).}
    \begin{ruledtabular}
    \begin{tabular}{lrrr}
     &\multicolumn{1}{c}{\textrm{$r_F$ raw}}&
    \multicolumn{1}{c}{\textrm{optical ac Stark}}&
    \multicolumn{1}{c}{\textrm{$r_F$ corrected}}\\
    \colrule
    $r_5$ & 5.382\,835\,18\,(75) & -3.56(36)[-7] & 5.382\,837\,10\,(78)\\
    $r_6$ & 1.537\,898\,38\,(21) & 8.59(98)[-9] & 1.537\,898\,36\,(21)\\
    $r_7$ & -0.863\,923\,40\,(12) & -1.42(14)[-6] & -0.863\,924\,62\,(17)\\
    $r_8$ & -2.463\,850\,49\,(37) & 0.35(25)[-7] & -2.463\,850\,41\,(37)\\
    $r_9$ & -3.582\,644\,30\,(50) & 5.02(75)[-8] & -3.582\,644\,12\,(50)\\
    \end{tabular}
    \end{ruledtabular}
\end{table}

Other systematic effects considered include the ac Zeeman shifts due to the ac magnetic field at the trap rf frequency, and microwave probe-induced shifts in $^3D_1$. Shifts from the trap-induced ac magnetic field depend only on the component perpendicular to the applied dc field~\cite{gan2018oscillating}. This is measured from an Autler-Townes splitting as described in previous work~\cite{arnold2020precision}.  The inferred field amplitude of $B_{\perp}=0.5368(11)\,\mu T$ implies a $\lesssim 1\times 10^{-8}$ fractional shift for all $r_F$ at the operating magnetic field. Microwave probe induced shifts arise from off-resonant microwave couplings to neighbouring magnetic substates. We determine the polarization components of the microwave field at the ion by measuring the coupling strengths on $\Delta m = (0,\pm 1)$ transitions at fixed microwave power. We estimate the probe-induced shift to be a $\sim 1.3\times 10^{-8}$ fractional correction for $^3D_1(F=6)$ Zeeman splitting. These shifts are both well below the stated uncertainties and omitted from \tref{tab:probestark}.


The $r_F$ results are summarized in \tref{tab:probestark} including corrections for 804 nm probe induced shift, the only significant systematic. To check repeatability, additional measurements were conducted for $F=5$ and $F=7$, which have the largest and smallest g-factor respectively. The repeated g-factor ratio measurement results corrected for systematics are $r_5=5.382\,835\,51\,(100)$ and $r_7=-0.863\,924\,64\,(20)$, which can be compared to the corresponding values in \tref{tab:probestark}. 

To determine $g_F$, we use 
$$\bar{g}_6=-0.071\,662\,506(33)$$
reported in \cite{zhiqiang2023176lu+}. The values of $g_F=\bar{g}_6 r_F$ are determined to be
\begin{align*}
g_5&=  -0.385\,747\,60\,(19), \\
g_6&=  -0.110\,209\,651\,(53), \\
g_7&=  0.061\,911\,003\,(31), \\
g_8&=  0.176\,565\,694\,(86), \\
g_9&=  0.256\,741\,26\,(12). 
\end{align*}
The uncertainties for all $g_F$ are limited by the uncertainty in $\bar{g}_6$ and not the uncertainties in the measured ratios.

\section{The residual quadrupole moment}
\label{Sect:analysis}
From theoretical calculations given in \cite{zhiqiang2020hyperfine}, first order corrections to g-factors can be written 
\begin{equation}
\delta g_F^{(1)}=\sum_{k,J'} C_{F,J,J'}^{k} \beta_{J,J'}^{(k)}
\end{equation}
where
\begin{align}
\beta_{J,J'}^{(1)} &=\frac
{\langle J\|\mathbf{m}\|J' \rangle
\langle J'\|T_1^{e}\|J \rangle}
{E_{J}-E_{J'}}\frac{\mu_I}{\mu_B I}, \\
\beta_{J,J'}^{(2)} &=\frac
{\langle J\|\mathbf{m}\|J' \rangle
\langle J'\|T_2^{e}\|J \rangle}
{E_{J}-E_{J'}}
\frac{Q}{2\mu_B I},
\end{align}
and
\begin{multline}
C_{F,J,J'}^{k}=2I\sqrt{\frac{2F+1}{F(F+1)}}
\begin{pmatrix}
I & k & I \\
-I & 0 & I\\
\end{pmatrix}^{-1} \\
\times
\begin{Bmatrix}
F & F & 1\\
J & J' & I
\end{Bmatrix}
\begin{Bmatrix}
F & J' & I\\
k & I & J
\end{Bmatrix}
\end{multline}
In these expressions, the nuclear magnetic dipole and electric quadrupole moments are defined as $\mu_I=\langle\mathbf{T}_k^n\rangle_I$ and $Q=2\langle\mathbf{T}_k^n\rangle_I$, where $\mathbf{T}_k^n$ and $\mathbf{T}_k^e$ are spherical tensor operators of rank $k$ that operate on the space of nuclear and electronic coordinates, respectively.  Thus, the g-factors for $^3D_2$ may be written
\begin{subequations}
\label{Eq:gfactor}
\begin{align}
g_5 = &-\tfrac{1}{3} g_J + \tfrac{4}{3} g_I +\delta g_5^{(2)}\nonumber\\
 &+\tfrac{8}{63} \beta_{2,3}^{(1)} -\tfrac{16}{45} \beta_{2,S}^{(1)}-\tfrac{68}{273}\sqrt{\tfrac{2}{3}} \beta_{2,3}^{(2)} + \tfrac{272}{195\sqrt{21}} \beta_{2,S}^{(2)}\\
g_6 = &-\tfrac{2}{21} g_J + \tfrac{23}{21} g_I +\delta g_6^{(2)} \nonumber\\
 &+\tfrac{374}{2205} \beta_{2,3}^{(1)} -\tfrac{4}{63} \beta_{2,S}^{(1)} +\tfrac{8}{35}\beta_{2,1}^{(1)}\nonumber\\
 &-\tfrac{2057}{9555}\sqrt{\tfrac{2}{3}} \beta_{2,3}^{(2)} -\tfrac{68}{1365 \sqrt{21}} \beta_{2,S}^{(2)} -\tfrac{136}{455}\beta_{2,1}^{(2)}\\
 g_7= &~\tfrac{3}{56} g_J +\tfrac{53}{56} g_I+\delta g_7^{(2)}\nonumber \\
&+\tfrac{81}{490}\beta_{2,3}^{(1)} +\tfrac{3}{280}\beta_{2,S}^{(1)} +\tfrac{221}{840}\beta_{2,1}^{(1)} \nonumber\\
&-\tfrac{81 \sqrt{6}}{3185}\beta_{2,3}^{(2)} + \tfrac{209}{3640}\sqrt{\tfrac{3}{7}}\beta_{2,S}^{(2)} -\tfrac{17}{280}\beta_{2,1}^{(2)}\\
g_8= &~\tfrac{11}{72}g_J + \tfrac{61}{72}g_I +\delta g_8^{(2)}\nonumber\\
&+\tfrac{247}{1890}\beta_{2,3}^{(1)} -\tfrac{11}{216}\beta_{2,S}^{(1)} + \tfrac{7}{40}\beta_{2,1}^{(1)} \nonumber\\
&\hfill+\tfrac{19}{315} \sqrt{\tfrac{2}{3}}\beta_{2,3}^{(2)} +\tfrac{143}{360 \sqrt{21}}\beta_{2,S}^{(2)} + \tfrac{7}{40}\beta_{2,1}^{(2)}\\
g_9= &~\tfrac{2}{9}g_J + \tfrac{7}{9}g_I +\delta g_9^{(2)}\nonumber\\
&\hfill+\tfrac{2}{27}\beta_{2,3}^{(1)} -\tfrac{28}{135}\beta_{2,S}^{(1)} +\tfrac{1}{9}\sqrt{\tfrac{2}{3}}\beta_{2,3}^{(2)} -\tfrac{4}{45}\sqrt{\tfrac{7}{3}}\beta_{2,S}^{(2)},
\end{align}
\end{subequations}
where $\delta g_F^{(2)}$ denote higher order corrections, fine-structure levels $^3D_J$ are denoted by the value of $J$, and $^1D_2$ by the symbol $S$.  Neglecting $\delta g_F^{(2)}$, there are a total of eight parameters, but only four of these are independent.  To show this, we introduce the orthogonal transformations
\begin{subequations}
\begin{align}
\begin{pmatrix}
\beta_{1}^{(1)}\\[6pt]
\beta_{2}^{(1)}\\[6pt]
\beta_{3}^{(1)}\\
\end{pmatrix} &=
\begin{pmatrix}
 \tfrac{7}{\sqrt{102}} &\tfrac{7}{\sqrt{102}} & \tfrac{2}{\sqrt{102}} \\[6pt]
 -\tfrac{44}{\sqrt{7701}} & \tfrac{58}{\sqrt{7701}} & -\tfrac{49}{\sqrt{7701}} \\[6pt]
 -\tfrac{9}{\sqrt{302}} & \tfrac{5}{\sqrt{302}} & \tfrac{14}{\sqrt{302}} \\
\end{pmatrix}
\begin{pmatrix}
\beta_{2,1}^{(1)}\\[6pt]
\beta_{2,S}^{(1)}\\[6pt]
\beta_{2,3}^{(1)}\\
\end{pmatrix}, \\
\begin{pmatrix}
\beta_{1}^{(2)}\\[6pt]
\beta_{2}^{(2)}\\[6pt]
\beta_{3}^{(2)}\\
\end{pmatrix} &=
\begin{pmatrix}
 \sqrt{\tfrac{98}{143}} &\sqrt{\tfrac{42}{143}} &\sqrt{\tfrac{3}{143}} \\[6pt]
 \tfrac{23}{\sqrt{4862}} & -\tfrac{31\sqrt{7}}{\sqrt{14586}} & \tfrac{56}{\sqrt{7293}} \\[6pt]
 \sqrt{\tfrac{7}{34}} & -\tfrac{5}{\sqrt{102}} & -\sqrt{\tfrac{28}{51}} \\
\end{pmatrix}
\begin{pmatrix}
\beta_{2,1}^{(2)}\\[6pt]
\beta_{2,S}^{(2)}\\[6pt]
\beta_{2,3}^{(2)}\\
\end{pmatrix},
\end{align}
\label{Transform}
\end{subequations}
and use an orthogonal decomposition of the equations.  In vector format, the $g_J$ and $g_I$ terms are first written in the form $(g_J-g_I) \mathbf{v}_J+g_I\mathbf{1}$, where $\mathbf{1}$ is a vector of ones and $\mathbf{1}\cdot  \mathbf{v}_J=0$.  The $k=1$ terms are then projected onto $\mathbf{v}_J$ and $\mathbf{1}$, which leaves a remainder  $\beta_1^{(1)} \mathbf{v}_1$.  Similarly, the $k=2$ terms are projected onto $\mathbf{v}_J$, $\mathbf{1}$, and $\mathbf{v}_1$, leaving a remainder $\beta_1^{(2)} \mathbf{v}_2$.  This procedure gives
\begin{equation}
\mathbf{g}_F=\bar{g}_J\mathbf{v}_J+\bar{g}_I\mathbf{1}+\bar{\beta}_1^{(1)}\mathbf{v}_1+\beta_1^{(2)}\mathbf{v}_2,
\end{equation}
where
\begin{equation}
\begin{aligned}
\mathbf{v}_J&=\left( -\tfrac{1}{3}, -\tfrac{2}{21}, \tfrac{3}{56}, \tfrac{11}{72}, \tfrac{2}{9} \right)^T, \\
\mathbf{v}_1&=\sqrt{\tfrac{17}{6}}\begin{pmatrix}
 -\frac{430}{4973} \\[6pt]
 \frac{15592}{174055} \\[6pt]
 \frac{18581}{174055} \\[6pt]
 \frac{571}{24865} \\[6pt]
 -\frac{4624}{34811}  
\end{pmatrix},\;
\mathbf{v}_2 =\sqrt{\tfrac{22}{13}}
\begin{pmatrix}
\tfrac{309098}{4804765}\\[6pt]
-\tfrac{796793}{4804765}\\[6pt]
\tfrac{139422}{4804765} \\[6pt]
\tfrac{831102}{4804765}\\[6pt]
-\tfrac{482829}{4804765}
\end{pmatrix},
\end{aligned}
\end{equation}
with
\begin{subequations}
\label{Eq:Reduced}
\begin{multline}
\bar{g}_J  = g_J - g_I + \tfrac{808\,037}{522\,165} \sqrt{\tfrac{2}{51}} \,\beta_1^{(1)} +\tfrac{2}{15} \sqrt{\tfrac{151}{51}} \,\beta_2^{(1)} \\
   -\tfrac{1\,824\,232}{11\,313\,575} \sqrt{\tfrac{2}{143}}\,\beta_1^{(2)} + \tfrac{113\,005}{64\,649} \sqrt{\tfrac{34}{143}}\,\beta_2^{(2)},
\end{multline}
\begin{multline}
\bar{g}_I  = g_I + \tfrac{2}{15} \sqrt{\tfrac{2}{51}}\,\beta_1^{(1)} - \tfrac{2}{15} \sqrt{\tfrac{151}{51}}\,\beta_2^{(1)} \\
-\tfrac{24}{325} \sqrt{\tfrac{2}{143}}\,\beta_1^{(2)} -\tfrac{12}{65} \sqrt{\tfrac{34}{143}}\,\beta_2^{(2)},
\end{multline}
\begin{equation}
\bar{\beta}_1^{(1)} = \beta_1^{(1)}-\tfrac{18\,915\,574}{1\,784\,627}\sqrt{\tfrac{3}{2431}}\,\beta_1^{(2)}-\tfrac{21}{13}\sqrt{\tfrac{3}{143}}\,\beta_2^{(2)}.
\end{equation}
\end{subequations}
We would then interpret $\bar{g}_I$ as the nuclear g-factor modified by the hyperfine interaction.  Higher order terms would add corrections to each of these parameters and leave a final orthogonal projection $\beta \hat{\mathbf{v}}$, where we take $\|\hat{\mathbf{v}}\|=1$.  This construction thus separates the correction terms by their expected relative sizes.

From the measured g-factors, we find
\begin{subequations}
\begin{align}
\bar{g}_J &= 1.156\,553\,17(54),\\
\bar{g}_I &= -1.47859(15)\times 10^{-4},\\
\bar{\beta}_1^{(1)} &= 5.57096(93)\times 10^{-4},\\
\beta_1^{(2)} &= -1.2795(77)\times 10^{-5},\\
\beta&=1.431(19) \times 10^{-6}.
\label{betaEq}
\end{align}
\end{subequations}
Uncertainties take into account the correlation from the uncertainty in $\bar{g}_6$.  Although this dominates the uncertainties in $g_F$ and $\bar{g}_J$, it is the uncorrelated errors in the ratios, $r_F$, that are more significant for the remaining parameters.  

If we neglect $\beta$, we have only four independent quantities for which to determine eight parameters: $g_I$, $g_J$, and $\beta_{2,J'}^{(k)}$ for $J'=1,S,3$ and $k=1,2$.  To reduce the number of parameters and accommodate estimation of the RQM, we take $g_I=-2.435047(16)\times10^{-4}$ as reported in \cite{zhiqiang2020hyperfine} and make use of the fact that coupling to the singlet state is expected to be small.  We therefore take $\beta_{2,S}^{(k)}\approx 0$, and 
\begin{equation}
\sum_{J'} \beta_{1,J'}^{(2)}\approx \beta_{1,2}^{(2)}=-\beta_{2,1}^{(2)}.
\end{equation}
with $\sum_{J'} \beta_{1,J'}^{(2)} = 3.130(21)\times 10^{-5}$ taken from \cite{zhiqiang2020hyperfine}.  We can then solve for the remaining parameters.  This is most easily achieved directly from Eq.~\ref{Eq:gfactor}.  

To estimate uncertainties, we first note that estimates of $\beta_{1,S}^{(k)}$ reported in \cite{zhiqiang2020hyperfine} were only a few percent of $\beta_{1,2}^{(k)}$.  We therefore take $0.1 \beta_{2,1}^{(k)}$ as an uncertainty for $\beta_{2,S}^{(k)}$ and $0.1 \beta_{2,1}^{(2)}$ as an uncertainty for $\beta_{2,1}^{(2)}$. Higher-order correction terms are indicated by the parameter $\beta$ in Eq.~\ref{betaEq}, which is larger than anticipated from $^3D_1$ contributions alone \cite{zhiqiang2020hyperfine}.  Hence we cannot simply neglect $\delta g_F^{(2)}$.  We therefore take $\beta$ as a bound on each $\delta g_F^{(2)}$ effecively treating each as an error term with mean of zero and variance $\beta^2$, which is equivalent to increasing the measurement uncertainties of $r_F$ to give an estimate of $\beta$ in Eq.~\ref{betaEq} that is statistically consistent with zero.  We then obtain the estimates
\begin{subequations}
\begin{align}
g_J&=1.156180(20)\\
\beta_{2,1}^{(1)}&=8.4(1.6)\times 10^{-4}\\
\beta_{2,S}^{(1)}&=0.0(9)\times 10^{-4}\\
\beta_{2,3}^{(1)}&=-0.9(2.5)\times 10^{-4}\\
\beta_{2,1}^{(2)}&=-3.1(3)\times 10^{-5}\\
\beta_{2,S}^{(2)}&=0.0(3)\times 10^{-5}\\
\beta_{2,3}^{(2)}&=9.1(3.5)\times 10^{-5}
\end{align}
\end{subequations}
Note that the result of $\beta_{2,1}^{(1)} =  8.4(1.6)\times 10^{-4}$ is consistent with $\beta_{1,2}^{(1)} = -8.6574(29)\times 10^{-4}$ from the $^3D_1$ measurements.

Finally, the parameters $\beta_{J,J'}^{Q(k)}$ associated with hyperfine-mediated quadrupole corrections may be written
\begin{equation}
\label{betaScale}
    \beta_{J,J'}^{Q(k)}=\beta_{J,J'}^{(k)} \cdot\mu_B I\cdot\frac{\langle J\|\Theta^{(2)}\|J' \rangle}{\langle J\|\mathbf{m}\|J' \rangle}.
\end{equation}
Using this relationship and matrix elements in Table.~\ref{table:ME}, gives
\begin{equation}
    \langle \delta\Theta(^3D_2,F,m)\rangle_F = 1.59(38)\times 10^{-4}\ ea_0^{2},
\end{equation}
for the $^3D_2$ RQM.  The uncertainty takes into account the correlation in the g-factor uncertainties and an additional 10\% uncertainty with the use of Eq.~\ref{betaScale} for which the matrix elements are calculated from theory but are expected to be good to within a few percent \cite{zhiqiang2020hyperfine}.  The RQM is of similar size to the $^3D_1$ value of $-2.54\times 10^{-4}\ ea_0^{2}$ and results in a low $10^{-19}$ shift of the clock frequency for typical dc confinement fields.
\begin{table}[htbp]
\caption{The reduced matrix elements of the magnetic dipole and electric quadrupole operators used in this work.  Matrix elements of $\mathbf{m}$ and $\Theta^{(2)}$ are given in units of $\mu_B$ and $e a_0^2$, respectively, where $\mu_B$ is the Bohr magneton, $e$ the fundamental unit of charge, and $a_0$ the Bohr radius.  Values are taken from \cite{zhiqiang2020hyperfine,arnold2015prospects}. }
\label{table:ME}
\begin{ruledtabular}
\begin{tabular}{cccc}
 $\ket{e}$ & $\ket{g}$ & $\langle e \|\textbf{m}\| g \rangle$ & $\langle e \|\Theta^{(2)}\| g \rangle$ \\
 \midrule
$^3D_{2}$ & $^3D_{1}$ & 2.055 & 4.523  \\
$^3D_{3}$ & $^3D_{2}$ & 2.094 & 4.969  \\
$^1D_{2}$ & $^3D_{2}$ & 0.218 & 1.319 \\
\end{tabular}
\end{ruledtabular}
\end{table}

In summary, we have provided precision measurements of the Land\'{e} g-factors for the 5d6s \textsuperscript{3}D\textsubscript{2} hyperfine levels of \textsuperscript{176}Lu\textsuperscript{+}.  The measurements, together with an analysis of hyperfine-mediated correction terms, have allowed us to make an initial assessment of the RQM of the hyperfine-averaged $^1S_0\leftrightarrow{^3}D_2$ clock transition.  Although this transition is relatively unexplored, key systematics are now well understood.  Magnetic field shifts can be readily calculated with accuracies of a few percent \cite{gan2018oscillating}, and the upper state lifetime \cite{paez2016atomic} is sufficiently long to take advantage of state-of-the-art laser stabilities yet short enough to limit probe-induced ac Stark shifts \cite{hyperRamsey1,hyperRamsey2}.  The blackbody radiation (BBR) shift \cite{arnold2018blackbody} is competitive with other clock systems and is the smallest among ion-clock species with a negative differential scalar polarizability, which allows suppression of micromotion shifts at a practical rf trap drive \cite{MagicSr}.  The RQM is one of the few systematics that cannot be easily determined from ab initio calculations.  Albeit a small effect, there is no good reason to suggest it will not become an important part of clock assessment of the $^1S_0\leftrightarrow{^3}D_2$  transition as it has for the $^1S_0\leftrightarrow{^3}D_1$ transition \cite{zhiqiang2023176lu+}.
\begin{acknowledgments}
This research is supported by the National Research Foundation, Singapore and A*STAR under its Quantum Engineering Programme (NRF2021-QEP2-01-P03) and through the National Quantum Office, hosted in A*STAR, under its Centre for Quantum Technologies Funding Initiative (S24Q2d0009); and the Ministry of Education, Singapore under its Academic Research Fund Tier 2 (MOE-T2EP50120-0014).
\end{acknowledgments}

\bibliography{gfactor3d2}

\end{document}